\documentstyle[aps,preprint]{revtex}
\draft
\begin{document}

\title{Shell stabilizes nuclear instability: A new phenomenon}
\author{L. Satpathy and S.K. Patra}
\address{Institute of Physics, Bhubaneswar - 751 005, India}
\date{\today}
\maketitle

\begin{abstract}
Infinite nuclear matter mass model and the relativistic mean field
theory show strong evidence of new neutron magic numbers 100, 150,
164; proton magic number 78 and new islands of stability around
N=100, Z $\simeq$ 62; N=150, Z=78; and N= 164, Z $\simeq$ 90 in
the drip-line regions of nuclear chart. It is shown that shell
effect stabilizes here the instability due to nuclear force giving
rise to a new phenomenon; in contrast to the phenomena of fission
isomer and super heavy elements where Coulomb instability is
overcome by the same.
\end{abstract}


Advent of heavy-ion reactions and production of radioactive ion
beams has opened up the possibility for the synthesis of exotic
nuclear species in the drip-line regions and the exploration of
new phenomena in nuclear physics. A prominent question of
fundamental importance is, whether classical magic numbers seen in
the valley of stability remain valid in the drip-line regions, or
new magic numbers manifest in such exotic nuclear terrains
characterized by high N/Z ratio. Such magic numbers in highly
unstable regions, could give rise to new islands of stability
driven by the shell effect which would have numerous consequences.
This possibility looks extremely attractive with the promise of a
new nuclear phenomenon, similar in importance to the phenomena of
double humped fission barrier with the manifestation of fission
isomer and the super heavy elements. While the latter two
phenomena occur in the normal nuclear terrain characterized by low
N/Z ratio appropriate for the valley stability, but vulnerable to
instability due to excessive Coulomb force originating from large
Z, the former pertains to a much larger N/Z ratio with the source
of instability in the repulsive part of the $n-n$ interaction
itself. This will be the first instance of stabilization by the
shell effect in the most unexpected domain of high instability of
nuclear force origin. This new class of phenomena is likely to be
observed as one moves across the valley of stability towards the
dripline, rather than along it. It envisages a nuclear landscape
without sharp coast line, but with a spread-out fuzzy region
dotted by tiny islands of exotic nuclear species. These are new
nuclear terrains in the drip-line regions unlike the super heavy
islands in the deep sea of instability. It is amusing to note that
the usual sea coast on the surface of our earth in many places is
quite fuzzed out with tiny islands. We provide strong evidence of
this exotic phenomenon from our study in infinite nuclear matter
(INM) mass model [1--3] and microscopic relativistic mean field
(RMF) theory [4], where we obtain new magic neutron numbers N=100,
150, 164 and proton magic number Z=78, and new islands of
stability N=100, Z $\simeq$ 62; N=150, Z=78; and N=164, Z $\simeq$
90 in the dripline regions. This is supported by our calculation
of shell correction energies using the usual Strutinsky
prescription[5]. Needless to say the possibility of occurrence of
such a phenomenon is expected on the general grounds of nuclear
dynamics.

The INM model is based [2] on the quantum mechanical infinite
nuclear matter, rather than the classical liquid used in the
liquid drop mass formulas of Bethe-Weiszacker type. Its main
elements are the INM ground-state and the generalized
Hugenholtz-Van Hove theorem [6]. A good account of this can be
seen in [2]. In this model the ground state energy $E^F(A,Z)$ of a
nucleus ($A, N, Z$) with asymmetry $\beta$  is considered
equivalent to the energy $E^s$ of a perfect sphere made up of
infinite nuclear matter at ground state with the same asymmetry
$\beta$ plus the residual characteristics energy $\eta$ called the
local energy. So,
\begin{equation}
E^{F} (A, Z) = E^{s}_{INM} (A, Z) + \eta (A, Z)
\end{equation}
with $E^{s}_{INM} (A, Z) = E(A, Z) + f (A, Z)$ where $f (A, Z)$
characterises the finite size effects given by
\begin{eqnarray}
f (A, Z) &=& a^{I}_{s} {A^{2/3}} + a^{I}_{c} (Z^2 -
5{{(3/16\pi)}^{2/3}} {Z^{4/3}}){A^{-1/3}} \\ \nonumber && - \delta
(A, Z).
\end{eqnarray}
The superscript $I$ stands for the INM characteristics of the
surface and Coulomb coefficients $a^I_s$ and $a^I_c$ respectively,
and $\delta(A,Z)$ is the usual pairing term, and `$F$' denotes the
quantities corresponding to finite nuclei.

Eq. (1) now becomes
\begin{equation}
E^{F} (A, Z) = E (A, Z) + f (A, Z) + \eta (A, Z).
\end{equation}
The term $E(A,Z)$ being the property of INM at the ground state,
will satisfy the generalized HVH theorem [6] of many-body theory,
\begin{equation}
\frac{E}{A} = [ ( 1+\beta ) {{\epsilon}_n} + ( 1-\beta )
{{\epsilon}_p} ],
\end{equation}
where, ${{\epsilon}_n} = {(\frac{\partial E}{\partial N}) \mid
}_Z$ and ${{\epsilon}_p} = {(\frac{\partial E}{\partial Z}) \mid
}_N$ are neutron and proton Fermi energies respectively. The
solution of Eq. (4) is of the form,
\begin{equation}
E = - a^{I}_v A + a^{I}_{\beta} {{\beta}^2} A
\end{equation}
where, $a^{I}_v$ and $a^{I}_{\beta}$ are identified as the volume
and asymmetry coefficients corresponding to INM.

Using Eqs. (3),(4) and (5) we arrive at three essential equations
of the model,
\begin{equation}
\frac{f}{A} - {\frac{N}{A}} {(\frac{\partial f}{\partial N}) \mid
}_Z
 - {\frac{Z}{A}} {(\frac{\partial f}{\partial Z}) \mid }_N
= \frac{E^F}{A} - {\frac{1}{2}} [ ( 1+\beta ) {{\epsilon}^{F}_n} +
( 1-\beta ) {{\epsilon}^{F}_p} ],
\end{equation}
\begin{equation}
- a^{I}_v + a^{I}_{\beta} {\beta ^2} = {\frac{1}{2}} [ ( 1+\beta )
{{\epsilon}^{F}_n} + ( 1-\beta ) {{\epsilon}^{F}_p} ]
 - {\frac{N}{A}} {(\frac{\partial f}{\partial N}) \mid }_Z
 - {\frac{Z}{A}} {(\frac{\partial f}{\partial Z}) \mid }_N,
\end{equation}
and
\begin{equation}
\frac{\eta (A, Z)}{A} =
  {\frac{1}{2}} [ ( 1+\beta ) {(\frac{\partial \eta}{\partial N}) \mid }_Z
+ {\frac{1}{2}} [ ( 1-\beta ) {(\frac{\partial \eta}{\partial Z})
\mid }_N].
\end{equation}

Eq. (6) determines $f$ through its fit to the combination of data
comprising total energy and neutron and proton separation
energies, which is subsequently used in Eq. (7) to obtain $a^I_v$
and $a^I_\beta$ , the properties of INM. Thus all the global
parameters are known. $\eta$ denotes the characteristic properties
of the nucleus which comprises shell, deformation and diffuseness
etc. and can be considered as its finger print. Once $E$ and $f$
are known, the empirical values of $\eta$ can be determined using
experimental binding energies in Eq. (3). Using the empirical
values of $\eta$ of all known nuclei, the $\eta$ of all unknown
nuclei are determined by using Eq. (8) for extrapolation. Over the
years, the success of this mass formula has been well demonstrated
[1--3,7--10]. Being built over a many-body theoretic foundation,
it has been shown to be particularly useful in extracting the
saturation properties and incompressibility of INM from nuclear
masses leading to the resolution of $r_0$-paradox [8,2]. Because
it uses the derivative of the energy in the form of the neutron
and proton separation energies in addition to the ground state
energies, it has unique ability for extrapolation which is re in
its prediction of shell quenching [9,10] for N=82, 126 shells.

The key quantity in the INM model is the local energy $\eta(A,Z)$
comprising all the characteristic features of a nucleus, which
includes predominantly shell effect and all other local effects
like deformation, diffuseness etc., and possibly unknown ones by
its construction. As shown by Nayak [10] and will be shown here
aposteriori, the plot of $\eta$ as a function of N for a given Z
called $\eta$-isoline shows a Gaussian structure with the peaks
lying at magic numbers. The height and width of the Gaussian is a
measure of the magicity/shell closure. Thus it has been shown that
$\eta$ carries a strong signature of nuclear shell-closure. The
two-neutron separation energy $S_{2n}$ is known to carry the
signature of shell closure in the valley of stability. The
$S_{2n}$ isolines show characteristic sharp bending just below and
above the magic numbers. Here we have made a thorough search of
the $\eta(A,Z)$, and $S_{2n}$ systematics obtained in our mass
predictions and found evidence of new shell-closures and new
islands of stability. The isolines for $S_{2n}$ and $\eta$ so
obtained are presented respectively in Figs. 1(a), 1(b) for the
region Z=50 $\sim$ 62, N=60 $\sim$ 110; and Fig 1(c), 1(d) for the
region Z=78-90, N=100-170 for even Z only to avoid clumsiness. It
may be noted that the range of isotopes chosen for presentation
here enables one to see in one sweep how the magicity evolves as
one moves from the valley of stability to the drip-line.

Expectedly in Fig. 1(b), the $\eta$ Gaussian for the double magic
nucleus Z=50, N=82 in the valley of stability shows the strongest
peak, and the corresponding ones for the other isotones
(Z=52$\sim$62) show progressively lowering of the height and
increase of width. The sharp shell-closure bending in $S_{2n}$
distributions (Fig. 1(a)) at N=82 are found to correlate with
those peaks supporting its magicity character. It is interesting
to find in Fig 1(b), the $\eta$ distributions for Z=58, 60 and 62,
show clear Gaussian peaks around N=100. The corresponding $S_{2n}$
isolines in Fig. 1(a) show shell-closure type bending for them
though not that prominently as at N=82, but nevertheless quite
conspicuously. Hence it is proper to recognise N=100 as a shell
closure. As for the protons, none of the three Gaussians is sharp
enough to qualify for a good shell closure, however they may be
regarded weakly magic. Hence we identify N=100, Z$\simeq$62 as a
new island of stability at the edge of the presently known
stability peninsula.

Similar analysis of Figs. 1(c) and 1(d) shows that $\eta$
distributions exhibit Gaussian peaks at N=126 for the values of Z
around 82 in the valley of stability which clearly correlate with
the characteristic bendings in the $S_{2n}$ isolines. As one moves
away towards the drip-line, one finds two more Gaussian
distributions around N=150 and N=164. The $\eta$ isolines for the
four elements, Z=84, 82, 80, 78 show well defined Gaussian peaks
with increasing sharpness and height around N=150. It is worth
noting here that the quality of the peak for Z=78 around N=150 is
as good and even better than the peak of Z=82 around N=126. From
Fig 1(c), one can see that the corresponding $S_{2n}$ isolines for
the four elements show shell closure type bending around N=150. So
we identify Z=78 as a good shell-closure and a new island of
stability around N=150, Z=78.

Moving further towards drip-line in Fig. 1(d), one finds a
Gaussian peak for the element Z=90 lying in the fringe of the
dripline region around N=164. Although it is of small height,
nevertheless it is quite conspicuous. We find the elements Z=88,
89 exhibiting peak structures also around the same neutron number
(Z=89 case not shown in figure). The corresponding $S_{2n}$
isolines in Fig. 1(c) show clear shell-closure type bending for
Z=88, 90 at N=164. Hence we identify a possible island of
stability around N=164, Z=90. It is worth recalling that N=164 has
been anticipated to be a magic number in many theoretical studies
in the literatures[11].

In a separate investigation we have carried out extensive study
microscopically in the RMF theory using NL3 [12] and NL-RA1[13]
interactions. The NL3 interaction has been widely used in recent
years in the calculation of varieties of nuclear properties like
binding energy, rms radii and giant resonances etc. and have been
accepted to be quite successful. In the present study, we expanded
the fields in harmonic oscillator basis taking $N_F = N_B =12$
major shells which is a reasonably large basis for the present
mass region[14]. We used a constant gap BCS pairing calculation to
take into account the pairing correlation. The pairing constant
gap is taken for the drip-line nuclei following the prescription
of Medland and Nix [15]. The formalism and calculation are quite
standard and have been widely used in the literature, the details
of which can be seen in refs. [14,16]. In our calculation we found
the results obtained with NL3 and NL-RA1 interactions, are quite
similar, therefore we present here the ones correspond to NL3 only
in Fig 2. In the upper, middle and the lower panel, we have
presented respectively the $S_{2n}(N,Z)$, and $S_{2n}(N,Z)$-
differential defined as $S_{2n}(N,Z)-S_{2n}(N+2,Z)$, and the
Hartree-BCS single particle spectra obtained in our calculations
for the relevant nuclei pertaining to the three new islands of
stability N=100, Z$\simeq$62; N=150 and Z=78; and N=164 and
Z$\simeq$90. In the upper panel, the solid lines connecting the
filled circles represent the cases where $S_{2n}$ show the
characteristic shell-closure bending at N=100, 150 and 164 for
nuclei with proton number around Z=62, 78 and 90 respectively in
remarkable agreement with the prediction of INM model. The
$S_{2n}$-differential, some time called shell-gap, is a more
sensitive quantity suitable for identification of shell closure.
These values shown in the middle panel corroborate the above
results more decisively through their prominent peaks. It is
heartening to see that these peaks occur at N=100, 150 and 164 as
found INM model. The Hartree-BCS single particle spectra shown in
the lower panel show conspicuous energy-gap for the nuclei with
relevant neutron and proton numbers. The energy-gap is a measure
of the goodness of shell-closure. In the lower panel the neutron
gaps at N=100, 150 and 164 persist when proton number is varied
between Z=56$\sim$62, Z=74$\sim$86, Z=88$\sim$94 respectively
showing the robustness of the magicity of these neutron numbers.
In combination with this fact, the energy-gaps seen in the proton
spectra at 62, 78 and 90 in $^{152}$Sm, $^{228}$Pt and $^{254}$Th
do support the identification of new islands of stability around
N=100, Z $\simeq$ 62; N=150, Z=78; and N=164, Z $\simeq$ 90.

A study of the $S_{2p}$-differential [$S_{2p}(N,Z)-S_{2p}(N,Z+2)$]
using the results of our RMF calculation was carried out for
N=150, 152 and 154 in the proton drip-line region. We find a well
defined peak for Z=78 showing its magicity in agreement with the
prediction of INM model.

By now it is well established that the shell correction energy
exhibits a pattern with inverted peaks at the position of the
shell-closures. To further ascertain the goodness of the magicity
of the neutron and proton numbers found above, we have calculated
the shell correction energies following the usual Strutinsky
prescription [5] using the single particle states obtained in our
RMF studies. This result presented in Fig. 3 shows the inverted
peaks occuring at N=100, 150, 164 for the regions around
Z$\approx$62, 78, 90 with energies $-4.3$ MeV, $-3.1$ MeV, $-2.9$
MeV, respectively. These values are expectedly lower than the
typical values of $-10$ MeV usually found for N=82 and 126 in the
valley of stability. However, they are substantially larger to
justify the magicity of these numbers.  Thus this calculation
shows that it is the shell effect of nuclear dynamics which gives
rise to the stability in a domain extremely vulnerable to
instability as a result of the repulsive part of the $n-n$ force.
Such stabilization manifests in the physical expansion of the
stability peninsula by pushing the drip-line away, and therefore
can be termed as a new phenomenon with potential for ample
experimental observations.  This predictions of the
semiphenomenological INM model being strongly corroborated
quantitatively by microscopic study in the RMF theory and further
reinforced by the calculation of shell effect is not only
satisfying, but also somewhat amazing. Thus we are constrained to
put adequate trust into this result.

In conclusion, the present studies carried out in both INM model
and microscopic RMF theory supplimented with the calculation of
shell correction energy, show strong evidence for the neutron
magic numbers 100, 150 and 164, proton magic number 78 in the
drip-line regions. The associated new islands of stability N=100,
Z $\simeq$ 62; N=150, Z $\simeq$ 78; and N=164, Z $\simeq$ 90 are
displayed by stars in the nuclear chart in Fig. 4. They lie in the
highly unstable regions closer to the drip-line. Thus our study
shows that the traditional magic numbers seen in the valley of
stability do not remain valid as one moves towards the drip-line
with progressively increasing N/Z ratio, and new magic numbers
make their appearance. It is exciting to note that the existence
of these new islands of stability in the highly unstable drip-line
regions, is due to the shell effect which stabilizes the
instability originating from the repulsive part of the nuclear
force. In this sense, it is a new nuclear phenomenon across the
valley of stability with high N/Z ratio, complementary to the
phenomena of fission isomers and super heavy elements observed
along the valley of stability with low N/Z ratio, where shell
effect stabilizes the instability due to repulsive Coulomb force.
This will give rise to new nuclear landscape in many regions
without sharp boundary, but with a fuzzed out coast dotted with
tiny islands.

\begin{figure}
  \caption{Isolines for the two-neutron
separation energy $S_{2n}$ and local energy $\eta$ obtained in the
infinite nuclear matter mass model.}
\end{figure}

\begin{figure}
  \caption{The two-neutron separation energy $S_{2n}$ , the
  $S_{2n}$-differential [$S_{2n}(N,Z)-S_{2n}(N+2,Z)$], and the
  neutron (solid line) and proton (dotted line) single-particle
  spectra for some of the nuclei, obtained in relativistic mean
  field theory with NL3 interaction are shown in the upper, middle
  and lower panel respectively.}
\end{figure}

\begin{figure}
  \caption{The shell correction energy for the nuclei around the new islands
of stability N=100, Z $\simeq$ 62; N=150, Z = 78; and N=164, Z
$\simeq$ 90.}
\end{figure}

\begin{figure}
  \caption{The new islands of stability N=100, Z $\simeq$ 62; N=150, Z = 78; and
N=164, Z $\simeq$ 90 are shown by stars in the nuclear chart. The
open squares represent the nuclei with experimentally known
binding energies ref. [17]. The neutron and proton drip-lines
shown as solid curves are the predictions of infinite nuclear
matter model.}
\end{figure}

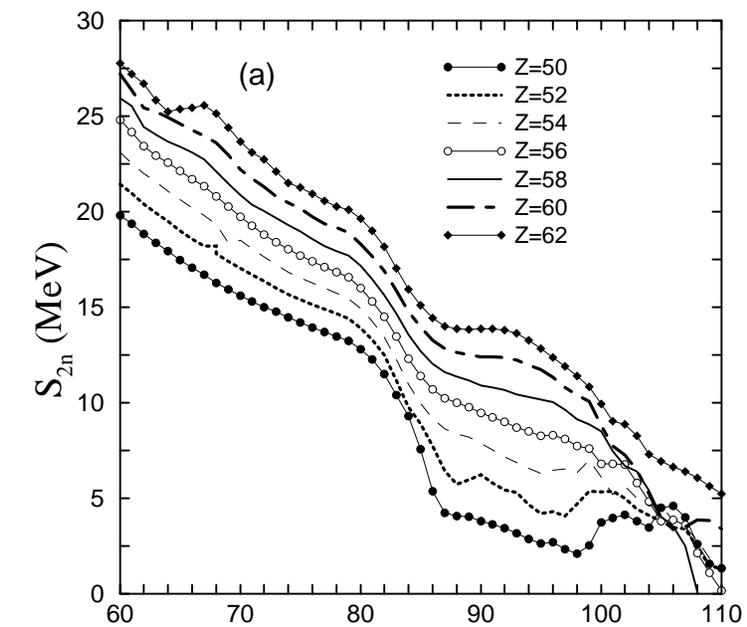
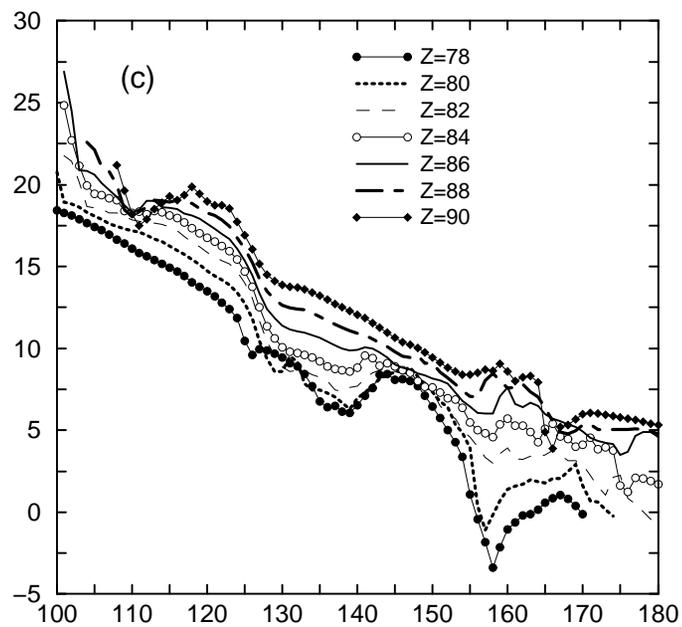
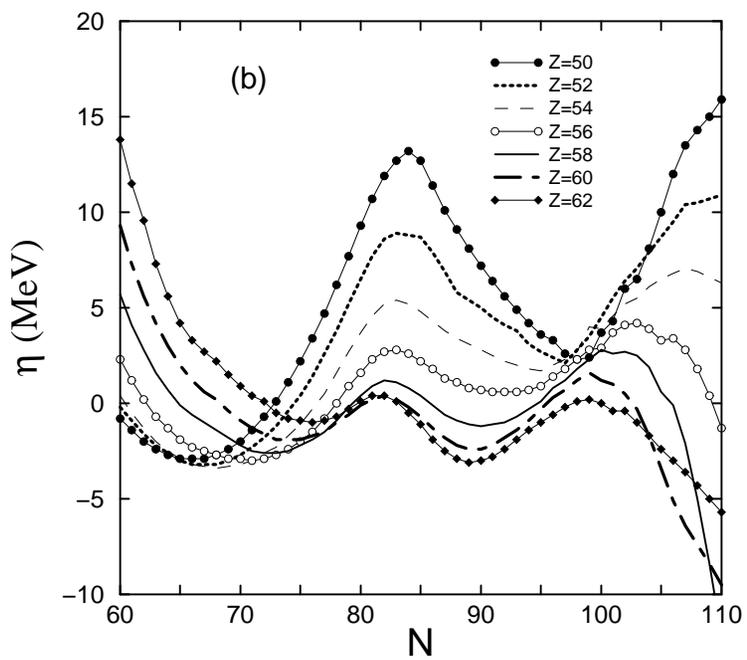
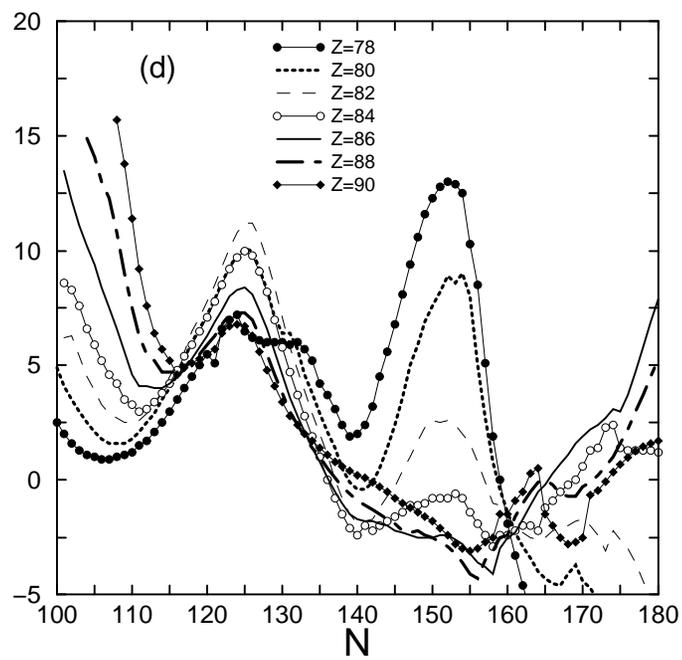
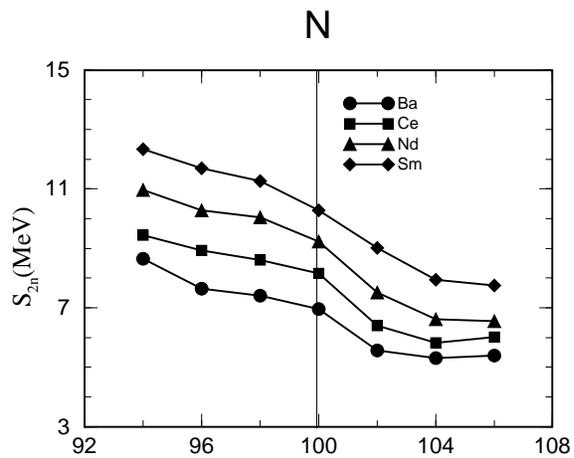
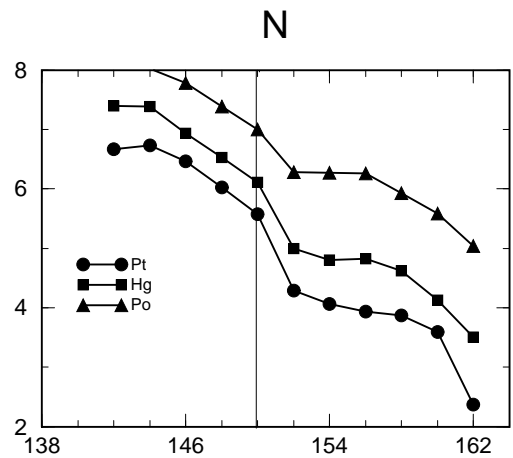
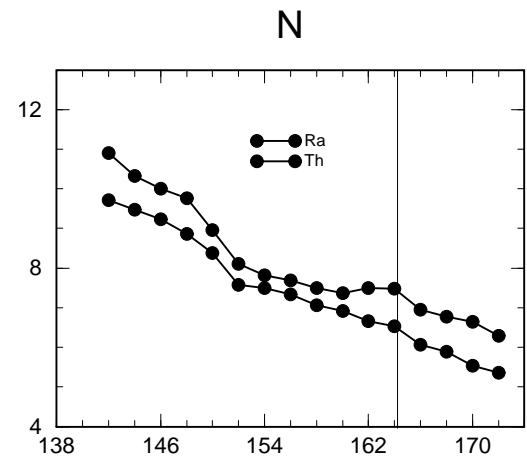
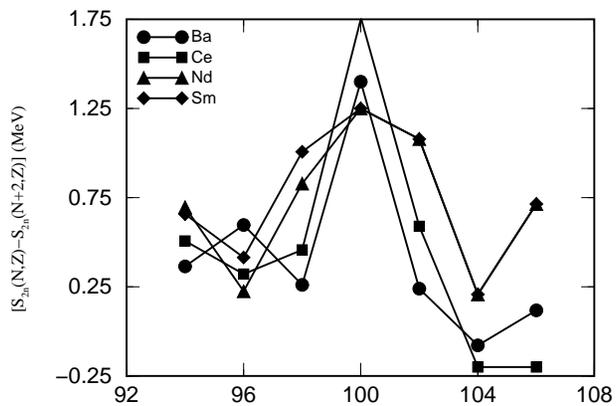
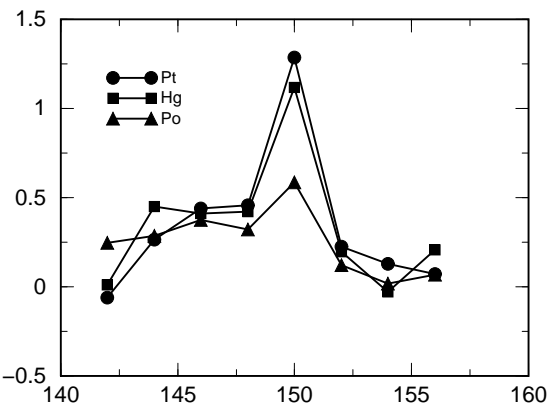
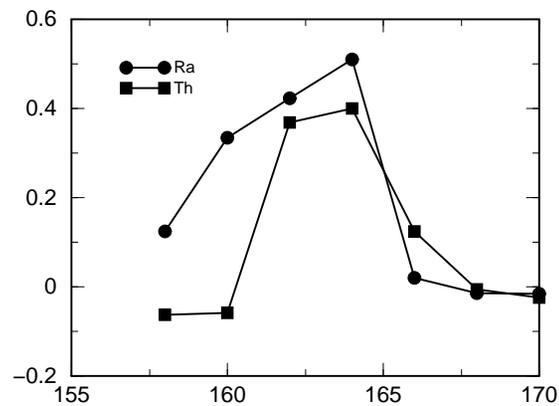
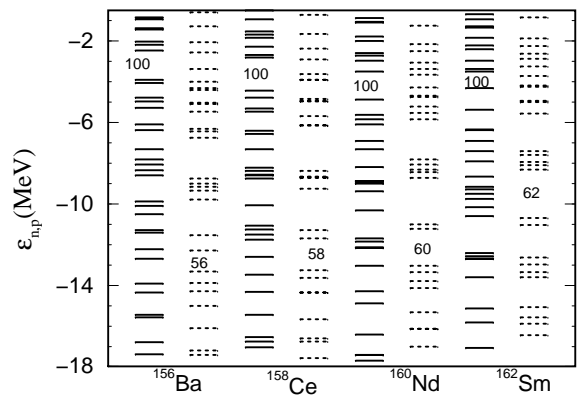
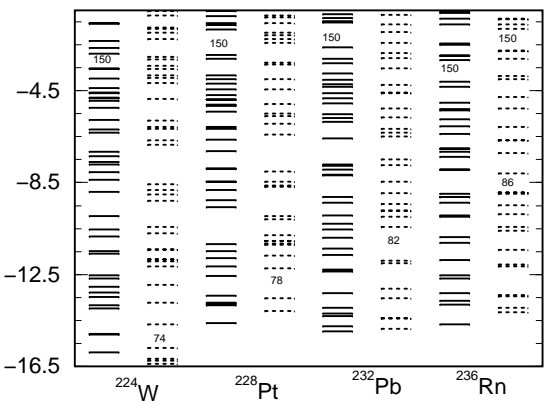
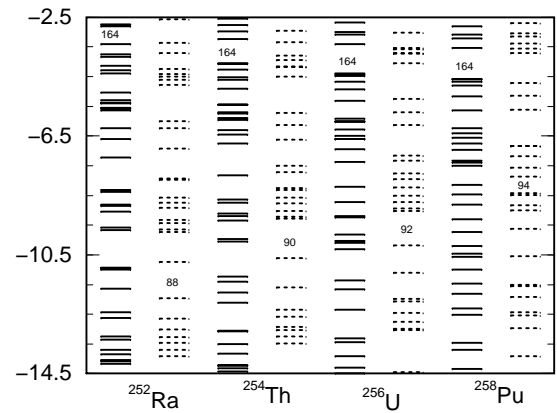
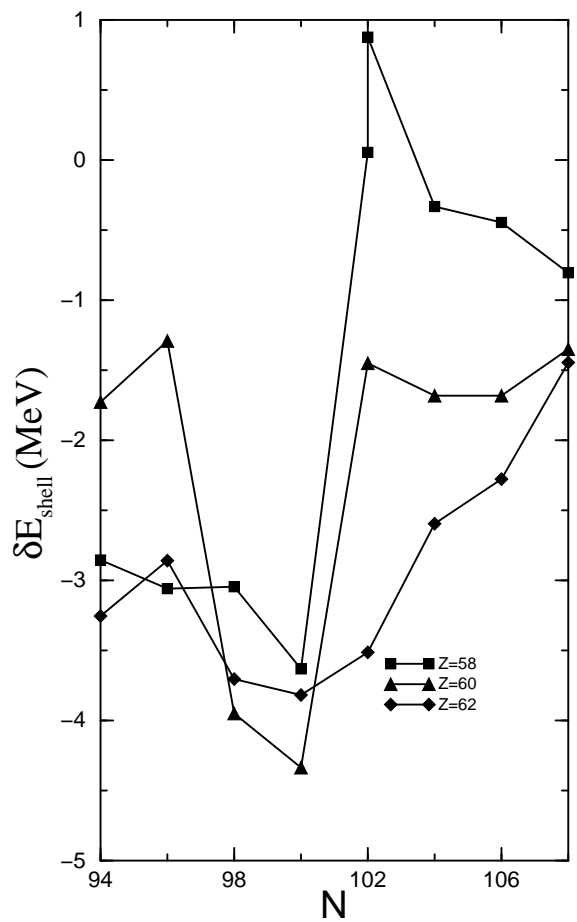
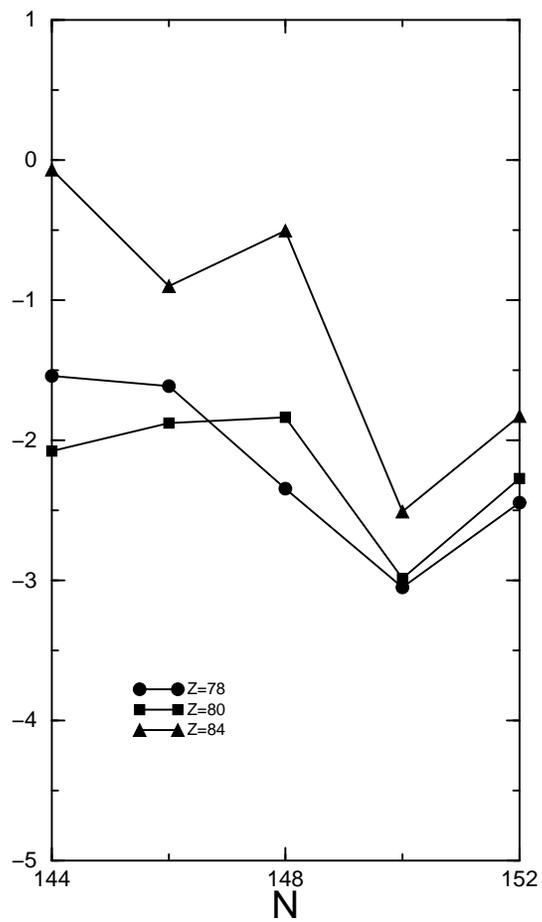
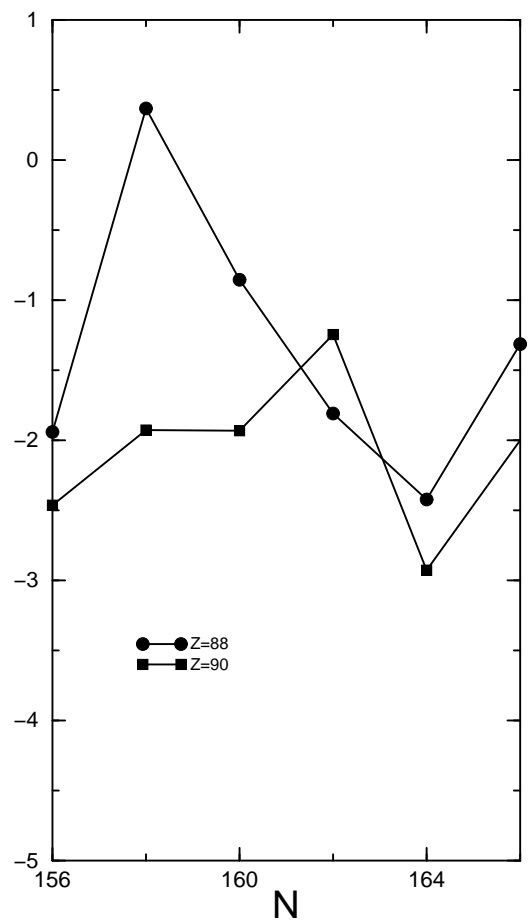
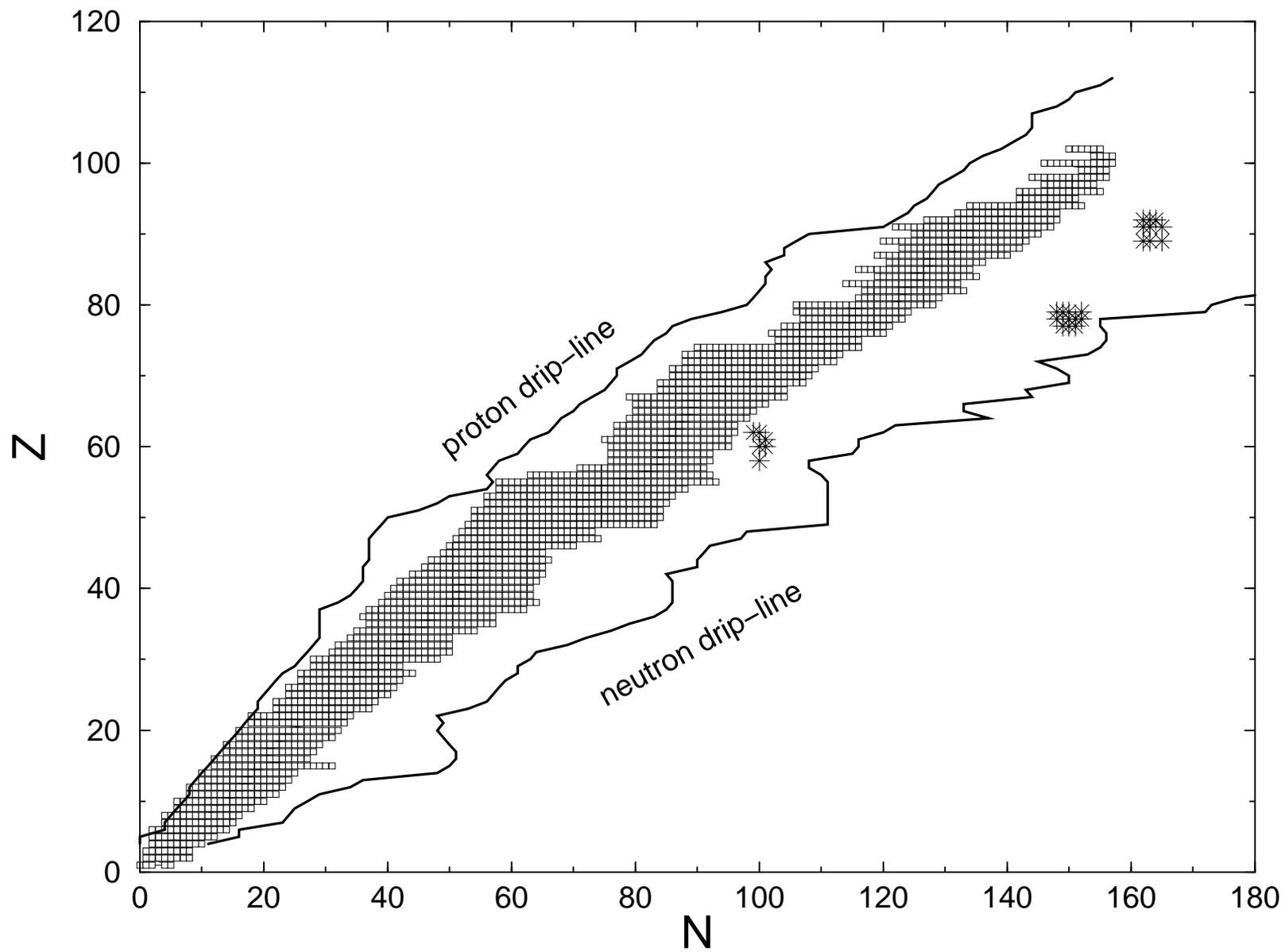
\end{document}